# Towards affective computing that works for everyone

Tessa Verhoef
*Leiden Institute of Advanced Computer Science*
*Leiden University*
Leiden, The Netherlands
t.verhoef@liacs.leidenuniv.nl
0000-0002-1219-3730

Eduard Fosch-Villaronga
*eLaw Center for Law and Digital Technologies*
*Leiden University*
Leiden, The Netherlands
e.fosch.villaronga@law.leidenuniv.nl
0000-0002-8325-5871

*Abstract*—Missing diversity, equity, and inclusion elements in affective computing datasets directly affect the accuracy and fairness of emotion recognition algorithms across different groups. A literature review reveals how affective computing systems may work differently for different groups due to, for instance, mental health conditions impacting facial expressions and speech or age-related changes in facial appearance and health. Our work analyzes existing affective computing datasets and highlights a disconcerting lack of diversity in current affective computing datasets regarding race, sex/gender, age, and (mental) health representation. By emphasizing the need for more inclusive sampling strategies and standardized documentation of demographic factors in datasets, this paper provides recommendations and calls for greater attention to inclusivity and consideration of societal consequences in affective computing research to promote ethical and accurate outcomes in this emerging field.

## I. INTRODUCTION

Diversity and inclusion are critical aspects of the responsible development of artificial intelligence (AI) technologies, including affective computing. Affective computing, which focuses on recognizing, interpreting, and responding to human emotions, has the potential to revolutionize various domains, such as healthcare, education, and human-machine interaction [1]. Capturing subjective states through technical means is challenging, though, and errors can occur, as seen with lie detectors not working adequately [2] or gender classifier systems misgendering users [3]. If used for ulterior decision-making processes, such inferences could have disastrous consequences for people, the impacts of which may vary depending on the context of an application, i.e., flagging innocent people as potential criminals in border control [4] or detrimentally affecting vulnerable groups in mental health care [5].

Given that single, unimodal data streams seem not robust enough to recognize human emotions, a growing trend in affective computing is using multimodal data to achieve machine interpretation, prediction, and understanding of human affective processes [6], [7]. These modalities include recognizing facial expressions, vocal tones, body posture and gestures, and other physiological signals such as heart rate and skin conductance [8]. Multimodal affective computing promises to measure user reactions to particular content better, create more interactive and engaging user experiences, and gain deeper insights into user behavior much more accurately than with unimodal systems [9].

However, combining different information strains is not straightforward [5], [10] and to what extent multimodal approaches will be able to solve existing problems in the field of affective computing remains an open question. Issues such as the existing disagreement on the nature and scientific understanding of emotions [11], or problems related to bias, discrimination, and injustice in affective computing [1], [12] likely remain. For instance, although emotion recognition algorithms have gained significant attention, they may have different outcomes for different groups due to health conditions, age, and gender [13], [14]. Mental health conditions such as depression or schizophrenia can affect facial expressions and speech, making it difficult to identify emotions accurately [15]. Non-neurotypical individuals may also have difficulty expressing emotions, while individuals with PTSD or phobias may display exaggerated or blunted emotional responses [16]. Concerning age, children's facial expressions and speech may be less distinct than adults and older adults' facial appearance changes due to aging can make it harder for algorithms to detect emotions accurately [17].

Given the growing interest in using these techniques in sensitive contexts such as healthcare and education, we explore how multimodal affective computing impacts diversity, equity, and inclusion in this article. In particular, we review the various ways human traits influence emotional expression and discuss its consequences for the current state of diversity and inclusion in affective computing. We then analyze an extensive list of datasets commonly used in affective computing and highlight how diverse and inclusive they are by juxtaposing them with the different grounds for discrimination that the law provides (i.e., religion or belief, origin, sexual orientation, sex, skin color, race, civil status, disability or chronic illness, or age), focusing on those characteristics that may affect emotion recognition the most. We anticipate that systems trained on the datasets currently available and used most widely may not work equally well for everyone and will likely have racial biases, biases against users with (mental) disabilities,

Supported by the Global Transformations and Governance Challenges Initiative at Leiden University for project Gendering Algorithms

and age biases because they derive from limited samples that do not fully represent societal diversity. We conclude by highlighting that there is still a long way to go for the field of affective computing to combat biases and inequalities that are typically exacerbated by the lack of diversity in datasets, technical teams, and the community [18]. Finally, we propose recommendations for improving diversity and inclusion in affective computing research.

## II. HUMAN TRAITS INFLUENCING EMOTIONAL EXPRESSION

Although physical and physiological markers to recognize emotions have gained significant attention in recent years, algorithms based on these markers may have different outcomes for different groups due to several factors that influence the recognition of emotions, such as age, health conditions and gender.

Emotion recognition algorithms may have limitations in detecting emotions in different age groups [19], [20]. For instance, children's facial expressions and speech may be less distinct than adults' [21] and may not have a full range of emotions, making it difficult to identify their emotional state accurately [22]. Furthermore, children may understand emotions differently than adults, and their expressions may not match their feelings, further complicating emotion recognition. One of the main challenges in emotion recognition algorithms for older adults is changes in facial appearance due to aging. As people age, their faces undergo various changes, such as the loss of muscle tone, wrinkle formation, and reduced eyebrow movements, making it harder for algorithms to detect emotions such as surprise or anger [23]. Also, if people undergo plastic surgery, which tends to happen later in life, this may affect emotion recognition systems [24]. In addition, health conditions that predominantly affect older individuals are known to affect emotional expressions. Alzheimer's Disease, for instance, is associated with impairments in the production of facial expressions and mood disorders [25], and Parkinson's Disease affects speech prosody and other communicative functions accompanied by an impact on mood [26].

Some mental health conditions affect facial expressions and speech, making it difficult to identify emotions accurately. For example, individuals with depression or schizophrenia often have a flattened affect, meaning they display fewer emotional expressions that, although not always affecting their subjective experience, make it challenging to detect emotional states [15], [27]. Similarly, non-neurotypical individuals may have difficulty expressing emotions, making it difficult to identify emotions accurately using physical markers [28]. Individuals with post-traumatic stress disorder (PTSD) or phobias may display exaggerated or blunted emotional responses, impacting emotion recognition [16].

Cultural differences may play a role as well [29], and emotion recognition algorithms may for example need different approaches for accurately detecting emotions in the deaf community [30]. Differences in body posture and facial expressions may carry linguistic meaning in sign languages [31] and may therefore be less reliable for detecting emotions in sign language users as compared to the hearing community.

Recent studies have also revealed that some systems based on facial features perform better in one gender than another, with generally lower accuracy for female faces [32]. This is not unlike well-known biases in face recognition, where for instance, algorithms developed by major tech companies were significantly less accurate in recognizing darker-skinned individuals, particularly women, than lighter-skinned individuals [33]. Several rationales may explain these disparities in accuracy and apparent errors in recognition performance. First, algorithms developed using non-diverse datasets may have gender biases in emotion recognition accuracy in underrepresented groups. Better emotion recognition was reported in female individuals, for instance, when females were overrepresented in the data [20]. Gender-balanced data does not guarantee balanced performance though, since other factors play a role as well. E.g. female faces have been found to be on average more similar to each other than male faces [32] and controlling for specific features known to be more prevalent in one gender over others, such as beards for males or make-up for females, balances performance more [32].

Affective computing researchers increasingly focus on the multimodal integration of multiple data sources as the solution for improving emotion recognition systems. However, the field of Machine Learning has identified problems that may arise when measurements from multiple datasets are combined, which may introduce increased "structured missingness" [34], referring to non-random patterns of missing data or underrepresentation. This problem is especially prevalent in data describing highly heterogeneous population characteristics, such as the expression of emotions. Besides problems in learning performance and prediction accuracy, structured missingness may perpetuate or exacerbate existing inequalities, especially when data from underrepresented groups is missing entirely.

For the field of affective computing, the extensive use of multimodal integration potentially involves combining datasets which are individually already relatively sparse and do not cover the range of genders, variety in mental health conditions, age groups, or other conditions that may have a direct effect on emotions. This can be very problematic and amplify biases that are already a major problem in single datasets. As an example, the most widely used facial expression dataset (Extended Cohn-Kanade (CK+) [35] has over 200 subjects in it, making it one of the most extensive datasets in terms of the number of subjects included for data collected in the lab, but it contains only two genders, not equally distributed (69% female), and a very skewed racial distribution with 81% euro-american, 13% afro-american, and 6% 'other.' The authors who released the original Cohn-Kanade dataset 23 years ago [36] already pointed out that many critical individual differences exist in facial expression features that vary with sex, age, and race. Moreover, they noted how various health conditions can affect facial expressions as well. They suggested including large samples of subjects with diverse backgrounds and health statuses to train emotion recognition systems that are robust to individual differences and work effectively for everyone. As we will show, the field has, unfortunately, yet to progress in this ideal direction.

If this probable inequality in access to functioning affective computing technology is not addressed, it's use in products and services will be problematic and potentially even illegal [18]. Given that religion or belief, origin, sexual orientation, sex, skin color, race, civil status, disability or chronic illness, or age are grounds for discrimination, errors in this area could lead to bias and discrimination, which is prohibited by law.

Although it is unsurprising that discriminatory outcomes may result from poor datasets that do not account for intersectional differences such as age, mental health, and gender, at this point, we do not know the magnitude of the problem, which is what we aim to unearth in this contribution.

## III. ANALYSIS OF DATASETS COMMONLY USED IN AFFECTIVE COMPUTING

To understand the magnitude of this issue, we analyze a diverse list of datasets commonly used in the field of affective computing. We base our selection on the most recent comprehensive review paper we could find [8], in which many datasets based on various signal modalities were listed.

An increasing number of datasets used in affective computing tasks contain (large amounts of) data collected from the web, such as written reviews on Amazon [37], [38] or IMDB [39] for textual sentiment analysis, or images and movies for bodily gesture and facial expression recognition through Google image search or YouTube [40], [41]. Since such datasets did not involve the recruitment of test subjects in a lab, no demographic information is available about the people performing the recorded emotions, making it impossible to assess our diversity dimensions for these sources. We, therefore, base our analysis on datasets that were created in the lab. In addition, we are aware of the fact that affective computing has been used to develop systems and applications targeted at specific populations, such as emotion detection and regulation systems for autism spectrum disorder (ASD) [42], Virtual Reality (VR) Therapy to help individuals with mental health problems, such as anxiety disorders or post-traumatic stress disorder (PTSD) [43], Affective Tutoring Systems for special needs education [44], or emotion-sensing Chatbots for mental health support [45]. Those systems are clearly meant to work for a specific target user group. However, here we focus on work published in general affect recognition aimed to be used ubiquitously by everyone, in general human-computer interaction, healthcare, social robotics, entertainment, advertising, automotive and education settings. In these application areas, it is essential that the technology is available and performs adequately for all potential users equally.

Considering all these criteria, we eventually included 26 datasets in our analyses with in total 1121 subjects. These datasets were released between 1998 and 2018 and span five different modalities: Speech, Face, Body, Physiological signals (e.g., EEG, GSR, temperature), and Multimodal (combinations of the first four). Table I lists all datasets with a short explanation of the type of collected data and a reference to the source paper, as well as analyzed demographic features, which are discussed in the results section.

## IV. RESULTS

Table I shows an overview of all datasets and their analyzed characteristics. We list which modality the data was based on, the year the dataset was released, the number of subjects included, the mean age of the subjects if it was mentioned in the paper, the percentage of female subjects included if specified, and the racial diversity of the subjects if mentioned. We analyzed the papers for the complete list of different grounds for discrimination that the law provides, but sexual orientation, religion, and civil status were never mentioned; therefore, we will not discuss them further in this paper.

### A. Race or cultural background: an incomplete task

When we look at the inclusion of subjects with different ethnic or cultural backgrounds, most papers actually do not mention it at all. Some explicitly state that all participants have the same background (indicated in the table as 'single'), while others include more diverse groups (indicated as 'multi'). For the multi-background datasets, we counted how many ethnic groups were represented out of four categories, Asian, Black, Latino and White. Fig. 1 plots these numbers for each modality and shows that this has been taken seriously mainly in the datasets for facial expression recognition, while datasets in the other modalities lack diversity. Most papers that describe datasets with diverse subject backgrounds also mention the exact composition [35], [46], [47], [52], [53], [55], [57]–[59], [61], [63]. For the pie chart in Fig. 2 all racial composition data from these studies were gathered, and we can see that some groups, especially Black and Latino, are highly underrepresented. Similar findings were highlighted by [20] for datasets based on facial features and by [71] for audiovisual datasets.

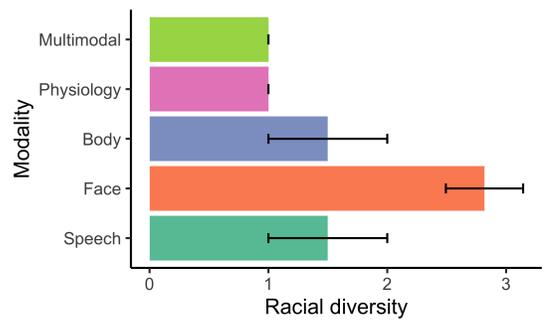

Fig. 1. Racial diversity (number of ethnic groups included), grouped by modality

### B. Sex and gender are different, but not always accounted for

When looking at the inclusion of different sexes, we see that most papers report the composition of their subject pool based on the percentage of male and female participants. Only one exception is observed (CASME II) where no information on this is mentioned. As also reported by [71], most datasets include equal or almost equal numbers of male and female subjects, besides a few exceptions where either female or male participants are outnumbered [35], [54], [61], [65] or not present at all [49], [52]. That said, this binary distinction may not work for contemporary societies in which other communities are not represented (intersex, transgender). Moreover, gender is different from sex and plays a crucial role in shaping one's self expression [3].

TABLE I
OVERVIEW OF DATASETS

| Modality | Database | Explanation | Year | # | Age | %Female | Race | Ref |
|---|---|---|---|---|---|---|---|---|
| **Speech** | Berlin Database of Emotional Speech (Emo-DB) | Audio data of sentences spoken in German by actors displaying various emotions. | 2005 | 10 | | 50.00% | | [46] |
| | Belfast Induced Natural Emotion (Belfast) | Describes three sets of data, with slight variations. Here we included the third and most diverse set. | 2012 | 60 | 26.35 | 50.00% | multi | [47] |
| **Body** | FAce and BOdy database (FABO) | Bimodal database with simultaneous video footage of both facial expressions and body gesture. | 2006 | 23 | 26.81 | 52.17% | | [48] |
| | THEATER Corpus | Movie clips coded with affective states and related to basic gestural form features. | 2009 | 2 | | 0.00% | | [49] |
| | GEneva Multimodal Emotion Portrayals (GEMEP) | Database of body postures and gestures collected from the perspectives of both an interlocutor and an observer. | 2010 | 10 | | 50.00% | | [50] |
| | EMILYA | Body gesture data captured with motion capture technology of actors expressing 8 emotions in 7 actions. | 2014 | 11 | 26 | 54.55% | | [51] |
| **Face** | JAFFE | Database with images of 7 facial expressions. | 1998 | 10 | | 100.00% | single | [52] |
| | Extended Cohn-Kanade (CK+) | Facial expression images of subjects who were instructed to perform 7 facial expressions. | 2010 | 210 | | 69.05% | multi | [35] |
| | MMI | Instead of a single image, this dataset consists of onset-apex-offset sequences. | 2010 | 25 | | 48.00% | multi | [53] |
| | Oulu-CASIA NIR-VIS (Oulu-CASIA) | Includes 2,880 image sequences captured with illumination invariant techniques, such as near-infrared (NIR). | 2011 | 80 | | 26.25% | multi | [54] |
| | BU 3D Facial Expression (BU-3DFE) | Contains 606 facial expression images in 3D captured from people with one of six facial expressions. | 2006 | 100 | | 60.00% | multi | [55] |
| | BU 4D Facial Expression (BU-4DFE) | Contains sequences of 3D images of facial expressions, adding an extra dimension. | 2008 | 41 | | 56.10% | multi | [56] |
| | BP4D-Spontaneous | Well-annotated 3D video database consisting of spontaneous facial expressions. | 2014 | 41 | 23.5 | 56.10% | multi | [57] |
| | 4DFAB | Dynamic 3D faces captured over a five-year period. | 2018 | 180 | 29.21 | 33.33% | multi | [58] |
| | Spontaneous Micro-expression (SMIC) | Contains 164 micro-expression video clips. | 2013 | 20 | | 30.00% | multi | [59] |
| | CASME II | Micro-expression videos. | 2014 | 35 | 22.03 | | | [60] |
| | Spontaneous Micro-Facial Movement (SAMM) | Another micro-expression video dataset with high resolution. | 2018 | 32 | 33.24 | 50.00% | multi | [61] |
| **Physiology** | DEAP | Combines physiological signals such as brain waves, skin conductance and body temperature. | 2012 | 32 | 26.9 | 50.00% | | [62] |
| | SEED | Contains brain wave data from subjects who were exposed to movie clips. | 2015 | 15 | 23.27 | 53.33% | single | [63] |
| | AMIGOS | Brainwaves, heartbeat and skin conductance were recorded using wearable sensors. | 2017 | 40 | 28.3 | 32.50% | | [64] |
| | Wearable Stress and Affect Detection (WESAD) | Combines data measured from multiple sensor modalities: blood volume pulse, heartbeat, electrodermal activity, electromyogram, respiration, body temperature, and threeaxis acceleration. | 2018 | 15 | 27.5 | 20.00% | | [65] |
| **Multimodal** | Interactive Emotional Dyadic Motion Capture (IEMOCAP) | Combines speech, body and facial signals through markers on the face, head, and hands, while actors were recorded during scripted and spontaneous spoken scenarios. | 2008 | 10 | | 50.00% | | [66] |
| | CreativeIT | Contains detailed full-body motion, visual-audio and text description data collected from actors during affective dyadic interactions. | 2011 | 16 | | 56.25% | | [67] |
| | MAHNOB-HCI | Combines video with physiological data using 6 video cameras, a head-worn microphone, an eye gaze tracker, and physiological sensors. | 2012 | 27 | 26.06 | 59.26% | | [68] |
| | RECOLA | Combines audio and video with physiological data recorded during spontaneous interactions between subjects. | 2013 | 46 | 22 | 58.70% | | [69] |
| | DECAF | Combines near-infra-red (NIR) facial videos with brain waves and various other physiological signals. | 2015 | 30 | 27.3 | 46.67% | | [70] |

*C. Prominent young age may disregard older groups*

For some, especially older dataset papers (N=7), no participants' age information was mentioned. Some papers reported the mean age of the subjects included in their datasets (N=14), sometimes including standard deviations (N=8), while others reported a range (N=14). In Table 1 we can see that the mean age of subjects is almost exclusively (with one exception) between 20 and 30 years of age across datasets. This is probably because many research groups use undergraduate or graduate students from their programs to participate in their studies. For the studies that reported an age range, Fig. 3 illustrates these for each dataset. We can see here that especially older age groups (50 and up) are highly underrepresented in the data. The one paper that includes a very large range (5-75, [58]), also reported percentages for different age groups. The very young and slightly older categories were still much less represented, with only 2.8% and 5.5%, respectively.

*D. (Mental) health or disability: an unfinished agenda*

The vast majority of papers introducing general-purpose affective computing datasets do not mention any inclusion of populations of subjects with varying (mental) health conditions. A few papers [62], [64], [68] mention explicitly that their participants were 'healthy', and these happen to all be datasets that include physiological data. One paper [65] was more specific and explicitly excluded participants with "pregnancy, heavy smoking, mental disorders, chronic and cardiovascular diseases." Another study [63] reported that they selected the subjects using the Eysenck Personality Questionnaire (EPQ), which characterizes personality in terms of Extraversion/Introversion, Neuroticism/ Stability and Psychoticism/Socialisation. The researchers noticed that their technology was less able to pick up on the physiological signals of introverted people or those with unstable mood; therefore, such individuals were deliberately excluded from participation.

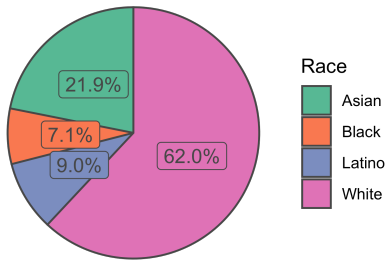

Fig. 2. Some groups are highly underrepresented

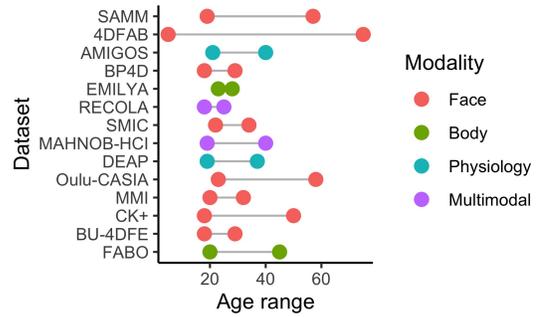

Fig. 3. Age ranges across datasets that mention age range

## V. DISCUSSION

Our findings shed light on the current state of inclusivity in affective computing datasets. The results reveal gaps in the representation of diverse populations in these datasets, particularly regarding race, age, and (mental) health/disability.

One noteworthy observation is that race is often not mentioned in the papers. When it is, most datasets lack diversity in terms of cultural backgrounds or origins. This is particularly evident in datasets for modalities other than facial expression recognition. This distinction is important because people do not only differ based on how they look but culture can also influence how people express emotions, for example, by emphasizing eye or mouth usage to identify people [29]. The underrepresentation of certain racial/ethnic groups, such as Black and Latino populations, is concerning and highlights the need for more inclusive sampling strategies to ensure these technologies work equally well across racial and cultural groups.

The age of participants is also an essential factor to consider in affective computing datasets. Mirroring findings for facial expressions datasets [20] and audiovisual datasets [71], the majority of datasets in our sample report a mean age between 20 and 30 years, indicating a bias towards younger age groups. This leads to an under-representation of older age groups, with limited data available for populations aged 50 and above. This is especially problematic since, as we reviewed, physical changes and age-related health conditions may significantly affect the expression of emotions in older populations.

Furthermore, the inclusion of populations with varying (mental) health conditions is lacking in affective computing datasets. Many papers do not mention any specific inclusion or exclusion criteria related to (mental) health or disability, and some even explicitly mention that their participants were "healthy." This lack of diversity in terms of mental health or disability status may limit the generalizability of affective computing technologies to populations with different mental health conditions and may perpetuate stigmas and biases related to mental health. We observed that especially for datasets based on physiological data, participants with health problems are sometimes explicitly excluded, which means that any multimodal dataset that includes physiological signals will have non-random missing values which can result in biased models that perform well for healthy individuals but poorly for those belonging to the underrepresented group. These findings highlight the need for researchers in the field of affective computing to be more mindful of inclusivity and actively consider the representation of diverse populations in their datasets to ensure that the technologies developed are more representative, equitable, and beneficial for all individuals.

Careful data collection with subjects in the lab is time-consuming and costly. It is, therefore, no surprise that recent datasets are often created by scraping data from the web. This has advantages, such as the large volume of data that can be collected in this way, which will also increase the inclusion of more diverse data sources. However, a disadvantage is that demographic information on the subjects in the dataset is not available, making it hard to measure and correct potential biases in the data. In this respect, emotion recognition algorithms also rely heavily on human annotations, which can be influenced by the annotators' demographic characteristics, and this can significantly impact the algorithms' accuracy [71].

To ensure consistent model performance for all target groups, sensitive applications such as emotion recognition must address representational bias in the data of both emotion expressers and annotators. Another way to potentially mitigate the adverse consequences of bias would be introducing a standardized (mandatory) way to document the inclusion of various relevant demographic factors in datasets. The recommendation proposed by [34] for machine learning in general, i.e., "appropriate sensitivity to social processes that underlie data generation and contextual awareness of potential social, cultural and historical determinants of discriminatory patterns are crucial for effective bias mitigation. Thus, involving experts with domain knowledge and social scientific training is vital;" applies to affective computing to a great extent. The mandatory Ethical Impact Statement in the papers presented at the Affective Computing and Intelligent Interaction (ACII) conference is a significant first step in this direction.

So far, we have focused mainly on the inclusion of diverse populations in datasets that are used for *training* affective computing systems, but it is equally crucial that systems are *tested* on diverse participants to make sure the recognition accuracy is generalizing and works equally well for diverse groups in society. Especially with the kinds of datasets that extensively use data downloaded from the internet, it is essential to assess potential biases by testing the technology on diverse users directly. Given the sometimes very sensitive application areas of affective computing, including the (mental) healthcare industry, it might not be excessive to apply similar guidelines surrounding diversity and inclusion used for clinical trials in medical sciences[1] to the testing of affecting computing technologies. It is good to remember that inferences based on subjective data could lead to disastrous consequences depending on the application context, where stakes could be extremely high. In other words, a recommender system that suggests a new song that you may like or a new movie to watch is not the same as a system that is meant to diagnose you with a particular disorder or disease [18] or a system that may be used in border control [4].

In conclusion, this study revealed that affective computing datasets generally lack diversity, with a limited representation of certain racial/ethnic groups and cultural backgrounds, sex and gender imbalances, skewed age demographics, and a total neglect of (mental) health/disability factors. This highlights the need for more inclusive sampling strategies and standardized documentation of demographic factors in datasets. Additionally, testing affective computing systems on diverse populations is crucial to ensure generalizability and accuracy. The sensitive nature of affective computing applications calls for guidelines similar to clinical trials in medical sciences. It is imperative to be mindful of the potential consequences of bias in subjective inferences, especially in such high-stakes contexts as those usually involved in affective computing.

ETHICAL IMPACT STATEMENT

This research addresses the ethical implications of inclusivity in affective computing datasets and its consequences for emotion recognition algorithms. This study underscores the need for greater diversity and representation in research samples by highlighting current datasets' limitations and potential biases in affective computing systems. The ethical implications of biased and inaccurate emotion recognition systems are significant, as they can impact vulnerable populations, including individuals with mental health conditions, children, older adults, and different genders. The potential consequences of bias in high-stakes contexts, such as decision-making and human-computer interactions, are also discussed, emphasizing the need for fair and accurate emotion recognition systems. The paper advocates for introducing inclusive sampling strategies, standardized documentation of demographic factors, and diversity and inclusion guidelines akin to those for clinical trials. The authors highlight that having more balanced datasets does not automatically lead to fair algorithms, and caution should be applied when considering their recommendations.

ACKNOWLEDGMENTS

We thank Joost Batenburg for providing support through the SAILS Program, a Leiden University wide AI initiative. This paper has also been partly funded by the Safe and Sound project, a project that received funding from the European Union's Horizon-ERC program (Grant Agreement No. 101076929).

---

[1] https://www.nimhd.nih.gov/resources/understanding-health-disparities/diversity-and-inclusion-in-clinical-trials.html